# The Oxford Olympics Study 2016:
# Cost and Cost Overrun at the Games

Bent Flyvbjerg[†]
Allison Stewart
Alexander Budzier

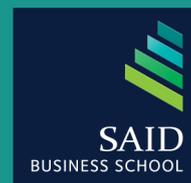
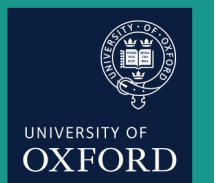

**Abstract**

Given that Olympic Games held over the past decade each have cost USD 8.9 billion on average, the size and financial risks of the Games warrant study. The objectives of the Oxford Olympics study are to (1) establish the actual outturn costs of previous Olympic Games in a manner where cost can consistently be compared across Games; (2) establish cost overruns for previous Games, i.e., the degree to which final outturn costs reflect projected budgets at the bid stage, again in a way that allows comparison across Games; (3) test whether the Olympic Games Knowledge Management Program has reduced cost risk for the Games, and, finally, (4) benchmark cost and cost overrun for the Rio 2016 Olympics against previous Games. The main contribution of the Oxford study is to establish a phenomenology of cost and cost overrun at the Olympics, which allows consistent and systematic comparison across Games. This has not been done before.

Main findings of the study are, first, that average actual outturn cost for Summer Games is USD 5.2 billion (2015 level), and USD 3.1 billion for Winter Games. The most costly Summer Games to date are London 2012 at USD 15 billion; the most costly Winter Games Sochi 2014 at USD 21.9 billion. The numbers cover the period 1960-2016 and include only sports-related costs, i.e., wider capital costs for general infrastructure, which are often larger than sports-related costs, have been excluded.

Second, at 156 percent in real terms, the Olympics have the highest average cost overrun of any type of megaproject. Moreover, cost overrun is found in all Games, without exception; for no other type of megaproject is this the case. 47 percent of Games have cost overruns above 100 percent. The largest cost overrun for Summer Games was found for Montreal 1976 at 720 percent, followed by Barcelona 1992 at 266 percent. For Winter Games the largest cost overrun was 324 percent for Lake Placid 1980, followed by Sochi 2014 at 289 percent.

Third, the Olympic Games Knowledge Management Program appears to be successful in reducing cost risk for the Games. The difference in cost overrun before (166 percent) and after (51 percent) the program began is statistically significant.

Fourth, and finally, the Rio 2016 Games, at a cost of USD 4.6 billion, appear to be on track to reverse the high expenditures of London 2012 and Sochi 2014 and deliver a Summer Games at the median cost for such Games. The cost overrun for Rio – at 51 percent in real terms, or USD 1.6 billion – is the same as the median cost overrun for other Games since 1999.

Given the above results, for a city and nation to decide to stage the Olympic Games is to decide to take on one of the most costly and financially most risky type of megaproject that exists, something that many cities and nations have learned to their peril.

## Why Study Cost and Cost Overrun at the Olympics?

Given that the six most recent Olympic Games, held over the decade 2004-2014, have cost on average USD 8.9 billion – not including road, rail, airport, and hotel infrastructure, which often cost more than the Games themselves – the financial size and risks of the Games warrant study. Furthermore, a focus on cost overruns as compared to the original budget is critical for future host cities to understand the implications of the investment they are undertaking. As part of bidding for hosting the Games, the International Olympic Committee (IOC) requires host cities and governments to guarantee that they will cover any cost overruns to the giant Olympics budgets. This means that the host city and nation are locked in to a non-negotiable commitment to cover any such increases. If overruns are likely this should clearly be taken into account in planning for the Games to get a realistic picture of the final outturn costs.

Moreover, given the current global economic climate and subsequent tightening of government spending in many countries, understanding the implications of major investments like the Games is critical for governments to make sound fiscal and economic decisions about their expenditures. For instance, cost overrun and associated debt from the Athens 2004 Games weakened the Greek economy and contributed to the country's deep financial and economic crises, beginning in 2007 and still playing out almost a decade later (Flyvbjerg 2011). Similarly, in June 2016 – less than two months before the Rio 2016 opening ceremony – Rio de Janeiro's governor declared a state of emergency to secure additional funding for the Games. When Rio decided to bid for the Olympics, the Brazilian economy was doing well. Now, almost a decade later, costs were escalating and the country was in its worst economic crisis since the 1930's with negative growth and a lack of funds to cover costs. Other countries – and especially those with small and weak economies – may want to make sure they do not end up in a similar situation by having a realistic picture of costs and risks of cost overruns before they bid for the Olympic Games. The data presented in this paper will allow such assessment.

Unfortunately, Olympics officials and hosts often misinform about the costs and cost overruns of the Games. For instance, in 2005 London secured the bid for the 2012 Summer Games with a cost estimate that two years later proved inadequate and was revised upwards with around 100 percent. Then, when it turned out that the final outturn costs were slightly below the revised budget, the organizers falsely, but very publicly, claimed that the London Games had come in under budget (BBC 2013). Such deliberate misinformation of the public about cost and cost overrun treads a fine line between spin and outright lying. It is unethical, no doubt, but very common. We can therefore not count on organizers and governments to provide us with reliable information about the real costs and cost overruns of the Olympic Games. Independent studies are needed, like the one presented here.

## Previous Studies of Cost and Cost Overrun

Interest in cost and cost overrun of the Games has been high since the establishment of the modern Olympics in 1896. As long ago as 1911 Baron Pierre de Coubertin, the man responsible for establishing the modern Games, referred to "the often exaggerated expenses incurred for the most recent Olympiads" (Coubertin, 1911), and in 1973 Jean Drapeau, the mayor of Montreal, infamously stated "The Montreal Olympics can no more have a deficit, than a man can have a baby," which caused some peculiar cartoons in Canadian media when the Montreal Games developed a large deficit due to outsized cost overruns (CBC 2006). Drapeau was wrong, and problems with cost and cost overrun are as prevalent today as they were in his time, and in Coubertin's before him.

Despite substantial interest in the cost of the Games, however, attempts to specifically and systematically evaluate such cost are few (Chappelet 2002, Essex and Chalkley 2004, Preuss 2004, Zimbalist 2015, Baade and Matheson 2016), while those that do attempt them are often focused on a specific Games (Bondonio and Campaniello 2006, Brunet 1995). Previous research on the cost of the Olympic Games has instead focused on whether the Games present a financially viable investment from the perspective of cost-benefit analysis. However, what to measure when determining the costs and benefits of the Games to a host country is open to debate and has varied widely between studies making it difficult to compare results across games. In particular, legacy benefits described in the bid are often intangible, and as such pose a difficulty in ex-post evaluations. The benefits of increased tourism revenue, jobs created by Olympic needs, or national pride are hugely varied and similarly difficult to quantify and compare. Costs are also hard to determine; for example, one could argue that if hotels in the host city have invested in renovations, and benefits of increased tourist revenues to those hotels are included in the analysis, then these costs should also be included in any accounting. Finally, the percentage of work that an employee in an outlying city spends on Games-related work would be exceptionally difficult to estimate.

Preuss (2004) contains the perhaps most comprehensive multi-Games economic analysis to date, looking at the final costs and revenues of the Summer Olympics from 1972 to 2008. Preuss finds that since 1972 every Organizing Committee of the Olympic Games (OCOG), which leads the planning of the Games in the host city, has produced a positive benefit as compared to cost, but only when investments are removed from OCOG budgets. This restricts the analysis narrowly to only OCOG activities, which typically represent a fairly small portion of the overall Olympic cost and therefore also denote too limited a view for true cost-benefit analysis. Further, other authors disagree with Preuss' findings, and have suggested that the net economic benefits of the Games are negligible at best, and are rarely offset by either revenue or increases in tourism and business (Malfas, Theodoraki, and Houlihan 2004). Furthermore, none of these studies have compared projected cost to final cost, which is a problem, because evidence from other types of megaprojects show that cost overruns may, and often do, singlehandedly cause positive projected net benefits to become negative (Flyvbjerg 2016; Ansar, Flyvbjerg, and Budzier forthcoming). The most recent study of the economics of the Olympics, published in *Journal of Economic Perspectives*, found that "the overwhelming conclusion is that in most cases the Olympics are a money-losing proposition for host cities" (Baade and Matheson 2016: 202).

In sum, we find for previous academic research on cost and cost overrun for the Olympic Games:

1. Earlier attempts to specifically and systematically evaluate cost and cost overrun in the Games are few;
2. Such attempts that exist are often focused on a specific Games or are small-sample research;
3. Earlier research on the cost of the Games has focused on cost-benefit analysis, with debatable delimitations of costs and benefits making it difficult to compare results across studies and Games.

Flyvbjerg and Stewart (2012) documented for the first time in a consistent and comparative fashion cost and cost overrun for a large number of Olympic Games. This study took its inspiration in comparative research more broadly looking at megaprojects and used a method for measuring cost and cost overrun that is the international standard in this research field. Flyvbjerg, Holm, and Buhl (2002), for example, provide an examination of rail, fixed-link (bridge and tunnel), and road projects, which finds that cost overruns are both prevalent and predictable, with average overruns of 45, 34, and 20 percent in real terms for each type of project, respectively. Their work has led to the development of a technique called "reference class forecasting" (Flyvbjerg 2008). Based on findings from behavioral economics, reference class forecasting develops budgets through a comparison with similar completed projects, rather than the bottom-up planning approach for each individual project that is commonly used. The reference class forecasting approach has been endorsed by the American Planning Association, and has been used in the UK, the Netherlands, Hong Kong, Denmark, and Switzerland, among others, to predict megaproject costs and benefits. Budzier and Flyvbjerg (2011) have confirmed similar results for major IT programs, as have Ansar et al. (2014) for large dams. Daniel Kahneman (2011: 251), Nobel Prize winner in economics, has called the use of reference class forecasting "the single most important piece of advice regarding how to increase accuracy in forecasting." Drawing on these insights, the present research expands and updates Flyvbjerg and Stewart (2012) in an attempt to further develop our understanding of cost and cost overrun in the Olympic Games.

## Measuring Cost and Cost Overrun for the Olympics

In the competition for hosting the Olympic Games cities pitch to the IOC and against each other their ideas for how to host the world's biggest sporting event and how to generate significant urban development in the process (Andranovich, Burbank, and Heying 2001). To demonstrate their ability to achieve these goals, bidding cities are required by the IOC to develop detailed plans in the form of so-called Candidature Files that are submitted to the Committee as part of the competition to host. The Candidature Files, or "bid books" as they are more commonly known, form part of the basis of the IOC's decision for the next host city.

One of the requirements for the bid book is that it includes a budget that details the expected investment by the host country's and city's governments in the Games, in addition to a budget for expected revenues (IOC 2004). In their bid book, governments of candidate cities and countries are also required by the IOC to provide guarantees to "ensure the financing of all major capital infrastructure investments required to deliver the Olympic Games" and "cover a potential economic shortfall of the OCOG" (*ibid*: p 93).

The Candidature File is a legally binding agreement, which states to citizens, decision makers, and the IOC how much it will cost to host the Games. As such the Candidature File represents the baseline from which future cost and cost overrun should be measured. If cost overrun later turns out to be zero, then decision makers made a well-informed decision in the sense that what they were told the Games would cost is what they actually cost, so they had the correct information to make their decision. If cost overrun is significantly higher than zero, then the decision was misinformed in the sense that it was based on an unrealistically low estimate of cost. However, such measurement of cost against a consistent and relevant baseline is rarely done. New budgets are typically developed after the Games were awarded, which are often substantially different to those presented at the bidding stage (Jennings 2012). These new budgets are used as new baselines, rendering measurement of cost overrun inconsistent and misleading both within and between Games. Using later baselines typically makes cost overruns look smaller and this is a strong incentive for using them, as in the case for London 2012 mentioned above. New budgets continue to evolve and develop over the course of the seven years of planning for the Games, until the final actual cost is perhaps presented, often several years after the Games' completion, if at all, as we will see.

In our effort to measure cost and cost overrun for the Games in a consistent and relevant manner, we searched for valid and reliable bid book and outturn cost data for both Summer and Winter Games, starting with the Rome 1960 Summer and Squaw Valley 1960 Winter Games, and continuing until the Sochi 2014 Winter and Rio 2016 Summer Games.

Costs for hosting the Games fall into the following three categories, established by the IOC:

1. *Operational costs* incurred by the Organising Committee for the purpose of "staging" the Games. The largest components of this budget are technology, transportation, workforce, and administration costs, while other costs include items like security, catering, ceremonies, and medical services. These may be considered the variable costs of staging the Games and are formally called "OCOG costs" by the IOC.

2. *Direct capital costs* incurred by the host city or country or private investors to build the competition venues, Olympic village(s), international broadcast center, and media and press center, which are required to host the Games. These are the direct capital costs of hosting the Games and are formally called "non-OCOG direct costs."

3. *Indirect capital costs* such as for road, rail, or airport infrastructure, or for hotel upgrades or other business investment incurred in preparation for the Games but not directly related to staging the Games. These are wider capital costs and are formally called "non-OCOG indirect costs."

The first two items constitute the sports-related costs of the Games and are covered in the present analysis. Non-OCOG indirect costs have been omitted, because (1) data on such costs are rare, (2) where data are available, their validity and reliability typically do not live up to the standards of academic research, and (3) even where valid and reliable data exist, they are typically less comparable across cities and nations than sports-related costs, because there is a much larger element of arbitrariness in what is included in indirect costs than in what is included in sports-related costs. It should be remembered, however, that the indirect costs are often higher than the direct costs. Baade and Matheson (2016: 205) found that for seven Games for which they could obtain data for both sports infrastructure and general infrastructure, in all cases was the cost of general infrastructure higher than the cost of sports infrastructure, sometimes several times higher.

Thus, our analysis compares each of OCOG costs and non-OCOG direct costs at two distinct points in time, bid budget and final outturn cost, for all Games since 1960 for which each of these four data points exist. This is 19 Games out of a total of 30 held between 1960 and 2016. For the remaining 11 Games, valid and reliable data have not been reported that would make it possible to establish cost overrun for these Games. This is an interesting research result in its own right, because it means – incredible as it may sound – that for more than a third of the Games between 1960 and 2016 no one seems to know what the cost overrun was. Such ignorance – willful or not – hampers learning regarding how to develop more reliable budgets for the Games. From a rational point of view, learning would appear to be a self-evident objective for billion-dollar events like the Games, but often that is not the case. For some Games, hiding costs and cost overruns seems to have been more important, for whatever reason.

Nevertheless, 19 out of 30 Games is 63 percent of all possible Games for the period under consideration (1960-2016), which we deem sufficient for producing interesting and relevant results. – We measured costs in both nominal and real (constant, not including inflation) terms, and in both local currencies and US dollars. We followed international convention and made all comparisons across time and geographies in real terms, to ensure that like is compared with like. Further detail on methodology and how costs were converted from nominal to real terms is available in the Appendix below.

## Costs of the Olympic Games 1960-2016

Table 1 shows actual outturn sports-related costs of the Olympic Games 1960-2016 together with the number of events and number of athletes in each Games.[1] Data on outturn cost were available for 25 out of the 30 Games 1960-2016. It should be mentioned that the Rio 2016 Summer Games had not yet been held at the time of writing the present paper. Preliminary data were therefore used for these Games.[2]

The difference between cost for the Summer and Winter Games is statistically significant ($W = 38$, $p = 0.02982$).[3] Costs for the Summer Games are significantly higher (median = USD 4.8 billion) than costs for the Winter Games (median = USD 2.0 billion). Costs for the two types of Games must therefore be considered separately in statistical terms.

---

[1] The Paralympic Games are not included here because they have only become fully integrated with the Olympic Games relatively recently and therefore do not compare across the period we study here.

[2] The latest available forecast of the Rio 2016 cost was made public on January 29, 2016. We have reviewed the figures and sources and deemed them reliable. The forecast comprises the operational cost managed by the Rio 2016 Organising Committee (BRL 7.4 billion) and the direct capital cost managed by the Autoridade Pública Olímpica (BRL 7.1 billion). As described above, the indirect capital cost, which are also managed by the Autoridade Pública Olímpica (BRL 24.6 billion), have been omitted from the comparisons in this paper for all Games.

[3] Significance is here defined in the conventional manner, with p≤0.05 being significant, p≤0.01 very significant, and p≤0.001 overwhelmingly significant. These analyses include the Rio 2016 forecasted cost.

*Table 1: Actual outturn sports-related costs of the Olympic Games 1960-2016, in 2015 USD*

| Games | Country | Type | Events | Athletes | Cost, billion USD |
|---|---|---|---|---|---|
| Rome 1960 | Italy | Summer | 150 | 5338 | n/a |
| Tokyo 1964 | Japan | Summer | 163 | 5152 | 0.282 |
| Mexico City 1968 | Mexico | Summer | 172 | 5516 | n/a** |
| Munich 1972 | Germany | Summer | 195 | 7234 | 1.009 |
| Montreal 1976 | Canada | Summer | 198 | 6048 | 6.093 |
| Moscow 1980 | Soviet Union | Summer | 203 | 5179 | 6.331 |
| Los Angeles 1984 | United States | Summer | 221 | 6829 | 0.719 |
| Seoul 1988 | South Korea | Summer | 237 | 8397 | n/a |
| Barcelona 1992 | Spain | Summer | 257 | 9356 | 9.687 |
| Atlanta 1996 | United States | Summer | 271 | 10318 | 4.143 |
| Sydney 2000 | Australia | Summer | 300 | 10651 | 5.026 |
| Athens 2004 | Greece | Summer | 301 | 10625 | 2.942 |
| Beijing 2008 | China | Summer | 302 | 10942 | 6.810 |
| London 2012 | United Kingdom | Summer | 302 | 10568 | 14.957 |
| Rio 2016* | Brazil | Summer | 306 | 10500 | 4.557 |
| *Average* | - | *Summer* | *239* | *8177* | *5.213* |
| *Median* | - | *Summer* | *237* | *8397* | *4.791* |
| Squaw Valley 1960 | United States | Winter | 27 | 665 | n/a |
| Innsbruck 1964 | Austria | Winter | 34 | 1091 | 0.022 |
| Grenoble 1968 | France | Winter | 35 | 1158 | 0.888 |
| Sapporo 1972 | Japan | Winter | 35 | 1006 | 0.117 |
| Innsbruck 1976 | Austria | Winter | 37 | 1123 | 0.118 |
| Lake Placid 1980 | United States | Winter | 38 | 1072 | 0.435 |
| Sarajevo 1984 | Yugoslavia | Winter | 39 | 1272 | n/a** |
| Calgary 1988 | Canada | Winter | 46 | 1432 | 1.109 |
| Albertville 1992 | France | Winter | 57 | 1801 | 1.997 |
| Lillehammer 1994 | Norway | Winter | 61 | 1737 | 2.228 |
| Nagano 1998 | Japan | Winter | 68 | 2176 | 2.227 |
| Salt Lake City 2002 | United States | Winter | 78 | 2399 | 2.520 |
| Torino 2006 | Italy | Winter | 84 | 2508 | 4.366 |
| Vancouver 2010 | Canada | Winter | 86 | 2566 | 2.540 |
| Sochi 2014 | Russia | Winter | 98 | 2780 | 21.890 |
| *Average* | - | *Winter* | *55* | *1652* | *3.112* |
| *Median* | - | *Winter* | *46* | *1432* | *1.997* |

Sources are listed under References. *) Current projected Rio 2016 figures have been used.
**) Mexican Peso and Yugoslavian dinar experienced hyperinflation during or after the Games.

We see that the most expensive Summer Games to date are London 2012 at USD 15.0 billion and Barcelona 1992 at USD 9.7 billion. For the Winter Games, Sochi 2014 is the most costly at USD 21.9 billion; Torino 2006 is the second-most costly at USD 4.4 billion. The least costly Summer Games are Tokyo 1964 at USD 282 million; the least costly Winter Games, Innsbruck 1964 at USD 22 million. It should again be remembered that wider capital costs (OCOG indirect costs) for urban and transportation infrastructure are not included in these numbers and that such costs are typically substantial.

Average cost for Summer Games 1960-2016 is USD 5.2 billion. Average cost for Winter Games over the same period is US 3.1 billion. The large difference between average and median cost for the Winter Games is mainly caused by the outlier of Sochi 2014 at USD 21.9 billion. Indeed, the Sochi 2014 Winter Olympics are the most costly Games ever, even when compared with the Summer Games. This is extraordinary, given the fact that cost for the Winter Games is typically much lower than for the Summer Games, with the median cost for Winter Games being less than half the median cost for Summer Games.

Figure 1 shows the development of cost 1960-2016. The trend lines indicate that the cost of the Games have increased over time. However, the apparent increase is statistically non-significant ($p = 0.764$, Winter Olympics; $p = 0.4473$, Summer Olympics). In statistical terms, therefore, we can argue for neither an increase or a decrease in cost over time.

*Figure 1: Time series of cost for Olympics 1960-2016*

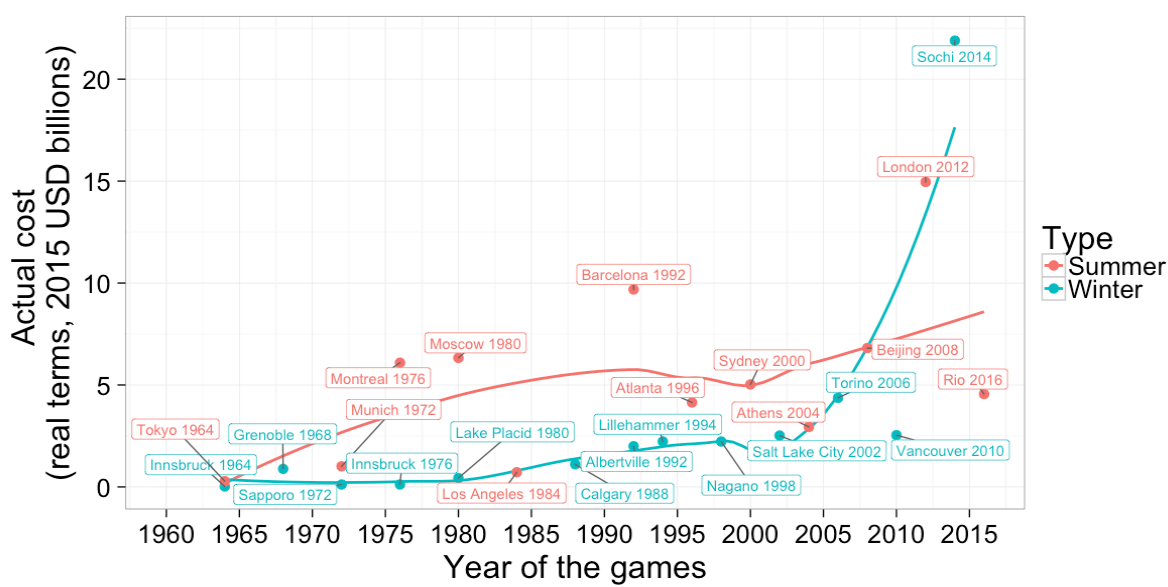

Table 2 shows cost per event and cost per athlete 1960-2016 in 2015 USD. These data were available for 25 of the 30 Games 1960-2016. The average cost per event for the Summer Games is USD 20 million (median USD 16 million). For the Winter Games it is USD 39.2 million (median USD 30 million). – The highest cost per event in the Summer Games was found for London 2012 at USD 50 million followed by Barcelona at USD 38 million. For the Winter Games, the highest cost per event was found for Sochi at USD 223 million followed by Torino 2006 at USD 52 million. Again we see that Sochi 2014 is an extreme outlier. The lowest cost per event was found for Tokyo 1964 at USD 1.7 million for the Summer games and Innsbruck 1964 at USD 0.6 million for the Winter Games.

*Table 2: Cost per event and cost per athlete in the Olympics 1960-2016, million 2015 USD*

| Games | Country | Type | Cost per event, mio. USD | Cost per athlete, mio. USD |
|---|---|---|---|---|
| Tokyo 1964 | Japan | Summer | 1.7 | 0.1 |
| Munich 1972 | Germany | Summer | 5.2 | 0.1 |
| Montreal 1976 | Canada | Summer | 30.8 | 1.0 |
| Moscow 1980 | Soviet Union | Summer | 31.2 | 1.2 |
| Los Angeles 1984 | United States | Summer | 3.3 | 0.1 |
| Barcelona 1992 | Spain | Summer | 37.7 | 1.0 |
| Atlanta 1996 | United States | Summer | 15.3 | 0.4 |
| Sydney 2000 | Australia | Summer | 16.8 | 0.5 |
| Athens 2004 | Greece | Summer | 9.8 | 0.3 |
| Beijing 2008 | China | Summer | 22.5 | 0.6 |
| London 2012 | United Kingdom | Summer | 49.5 | 1.4 |
| Rio 2016* | Brazil | Summer | 14.9 | 0.4 |
| *Average* | *-* | *Summer* | *19.9* | *0.6* |
| *Median* | *-* | *Summer* | *16.0* | *0.5* |
| Innsbruck 1964 | Austria | Winter | 0.6 | 0.02 |
| Grenoble 1968 | France | Winter | 25.4 | 0.8 |
| Sapporo 1972 | Japan | Winter | 3.4 | 0.1 |
| Innsbruck 1976 | Austria | Winter | 3.2 | 0.1 |
| Lake Placid 1980 | United States | Winter | 11.5 | 0.4 |
| Calgary 1988 | Canada | Winter | 24.1 | 0.8 |
| Albertville 1992 | France | Winter | 35.0 | 1.1 |
| Lillehammer 1994 | Norway | Winter | 36.5 | 1.3 |
| Nagano 1998 | Japan | Winter | 32.7 | 1.0 |
| Salt Lake City 2002 | United States | Winter | 32.3 | 1.1 |
| Torino 2006 | Italy | Winter | 52.0 | 1.7 |
| Vancouver 2010 | Canada | Winter | 29.5 | 1.0 |
| Sochi 2014 | Russia | Winter | 223.4 | 7.9 |
| *Average* | *-* | *Winter* | *39.2* | *1.3* |
| *Median* | *-* | *Winter* | *29.5* | *1.0* |

*Sources are listed under References. *) Projected final Rio 2016 costs have been used.*

For cost per athlete, we found the Winter Games to be twice as costly as the Summer Games. The average cost per athlete is USD 599,000 for the Summer Games (median USD 453,000) and USD 1.3 million for the Winter Games (median USD 990,000). However, the difference is statistically non-significant (W = 97, p = 0.3203). The highest cost per athlete in the Summer Games was found for London 2012 at USD 1.4 million, followed by Moscow 1980 at USD 1.2 million.[4] For the Winter Games, the highest cost per athlete was found for Sochi 2014 at USD 7.9 million and Torino 2006 at USD 1.7 million. The lowest cost per athlete in the Summer Games was found for Tokyo 1964 at USD 55,000, and in the Winter Games for Innsbruck 1964 at USD 20,000.

---

[4] The Moscow 1980 Summer Games were boycotted by 65 nations as a protest against the 1979 Soviet invasion of Afghanistan. The number of participating athletes was therefore lower than anticipated, driving up cost per athlete.

Figure 2 shows the correlation of cost per athlete with time. We see a shift in trend from cost per athlete being generally higher for the Summer than for the Winter Games until the mid 1980's, after which the Winter Games become more costly than the Summer Games, in terms of cost per athlete. We also see that cost per athlete was generally decreasing for the Summer Games from the mid-1980's until the early noughties, after which cost per athlete has been increasing for both the Summer and Winter Games, driven mainly by London 2012 and Sochi 2014. Overall, however, the changes over time are statistically non-significant for both Summer Games ($p = 0.4762$), Winter Games ($p = 0.1523$), and all Games ($p = 0.5399$).

*Figure 2: Time series of cost per athlete for Olympics 1960-2016, with and without outlier*

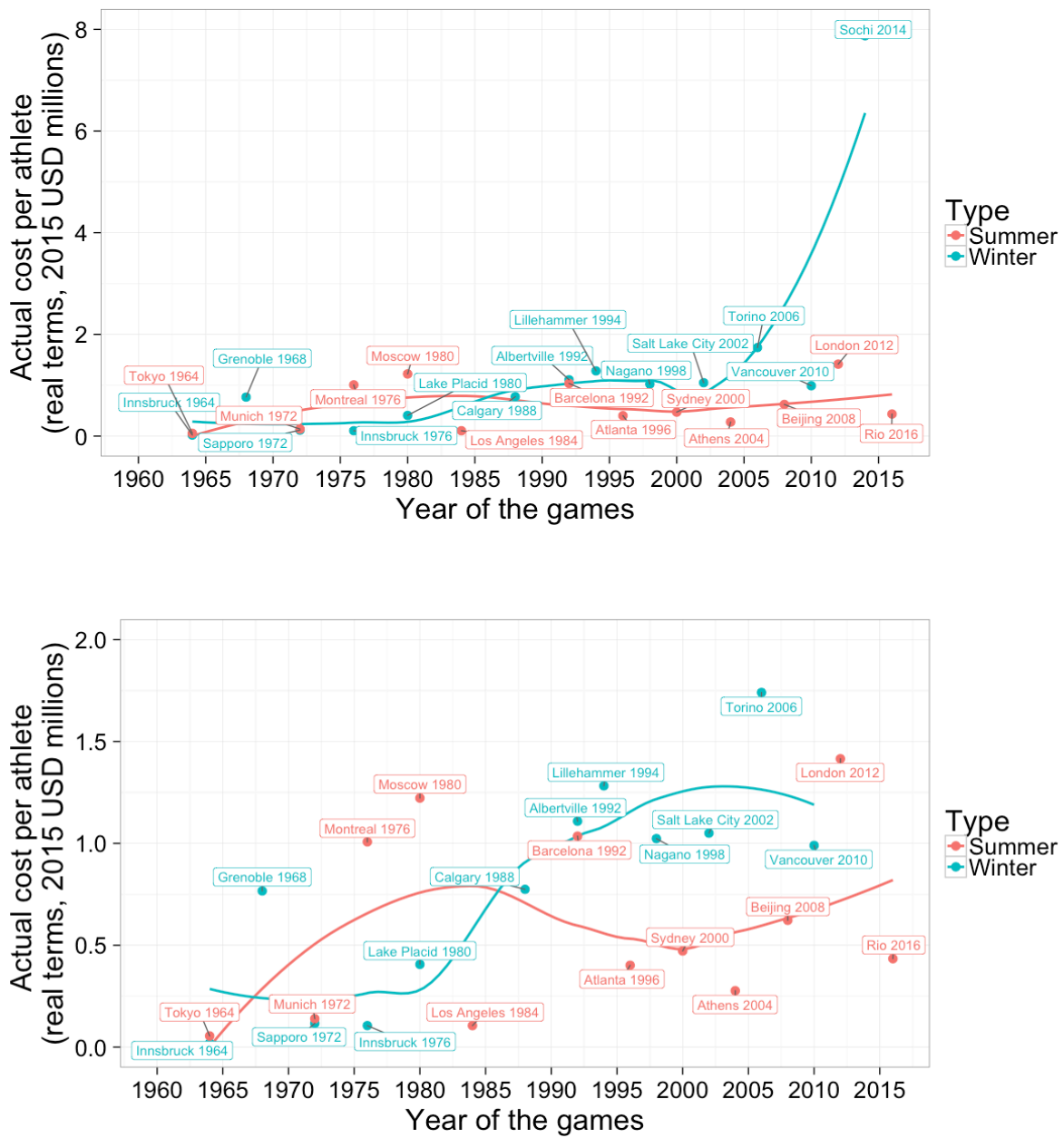

**Cost Overrun at the Games**

Table 3 shows percentage cost overrun in real terms for the Olympic Games 1960-2016. Data on cost overrun were available for 19 of the 30 Games 1960-2016. Statistical tests of the difference between bid budgets and final costs show this difference to be statistically overwhelmingly significant ($V = 190$, $p < 0.0001$). That is to say, cost overruns are overwhelmingly manifest for the Olympics, in statistical terms. – It should be mentioned that if the cost overruns had been calculated in nominal terms (including inflation) they would have been significantly larger. In this sense the numbers shown are conservative.

*Table 3: Sports-related cost overruns, Olympics 1960-2016; calculated in local currencies, real terms*

| Games | Country | Type | Cost overrun % |
|---|---|---|---|
| Montreal 1976 | Canada | Summer | 720 |
| Barcelona 1992 | Spain | Summer | 266 |
| Atlanta 1996 | United States | Summer | 151 |
| Sydney 2000 | Australia | Summer | 90 |
| Athens 2004 | Greece | Summer | 49 |
| Beijing 2008 | China | Summer | 2 |
| London 2012 | United Kingdom | Summer | 76 |
| Rio 2016* | Brazil | Summer | 51 |
| *Average* | - | *Summer* | *176* |
| *Median* | - | *Summer* | *83* |
| Grenoble 1968 | France | Winter | 181 |
| Lake Placid 1980 | United States | Winter | 324 |
| Sarajevo 1984 | Yugoslavia | Winter | 118 |
| Calgary 1988 | Canada | Winter | 65 |
| Albertville 1992 | France | Winter | 137 |
| Lillehammer 1994 | Norway | Winter | 277 |
| Nagano 1998 | Japan | Winter | 56 |
| Salt Lake City 2002 | United States | Winter | 24 |
| Torino 2006 | Italy | Winter | 80 |
| Vancouver 2010 | Canada | Winter | 13 |
| Sochi 2014 | Russia | Winter | 289 |
| *Average* | - | *Winter* | *142* |
| *Median* | - | *Winter* | *118* |

*Sources are listed under References.  *) Projected final Rio 2016 costs have been used*

We find the following averages and medians for cost overrun in real terms:

- All Games: average cost overrun is 156 percent (median 90 percent).
- Summer Games: average cost overrun is 176 percent (median 83 percent).
- Winter Games: average cost overrun is 142 percent (median 118 percent).

Even though the difference between average cost overrun for the Summer and Winter Games is relatively large at 34 percentage points, the difference between the two types of Games in terms of cost overrun is statistically non-significant (W = 49, p = 0.7168). In statistical terms there is therefore no difference between cost overrun in the Summer and Winter Games and the data may be pooled for statistical analyses, for instance in making more accurate reference class forecasts of budgets for future Olympic Games (Flyvbjerg 2008).

We further see that:

- 15 of 19 Games (79 percent) have cost overruns above 50 percent.
- 9 of 19 Games (47 percent) have cost overruns above 100 percent.

Judging from these statistics it is clear that large risks of large cost overruns are inherent to the Olympic Games.

For the Summer Games, the largest cost overrun was found for Montreal 1976 at 720 percent, followed by Barcelona 1992 at 266 percent. The smallest cost overrun for the Summer Games was found for Beijing 2008 at two percent, followed by Athens 2004 at 49 percent. For the Winter Games, the largest cost overruns are Lake Placid 1980 at 324 percent followed by Sochi 2014 at 289 percent. The smallest cost overrun for the Winter Games was found for Vancouver 2010 at 13 percent, followed by Salt Lake City 2002 at 24 percent.

The vigilant reader may be skeptical that the lowest cost overrun of all Games was found for Beijing 2008 at two percent. China is known for its lack of reliability in economic reporting (Koch-Weser 2013). However, the total cost of USD 6.8 billion and the cost per athlete of USD 622,000 for the Beijing 2008 Games are both higher than for the majority of other Summer Games (see Tables 1 and 2). The reported costs are therefore deemed adequate for hosting the Beijing Games and we have seen no direct evidence that the official numbers have been manipulated and should be rejected for this reason. Like other observers of economic data from China we therefore include the numbers, with the caveat that they are possibly less reliable than numbers from other nations, given China's history of doctoring data. Again, this means that our averages for cost overrun in the Olympic Games are conservative.

We further observe about cost overrun in the Olympic Games, based on the data presented above:

1. *All Games, without exception, have cost overrun*. For no other type of megaproject is this the case. For other project types, typically 10-20 percent of projects come in on or under budget. For the Olympics it is zero percent. It is worth considering this point carefully. A budget is typically established as the maximum – or, alternatively, the expected – value to be spent on a project. However, in the Games the budget is more like a fictitious minimum that is consistently overspent. Further, even more than in other megaprojects, each budget is established with a legal requirement for the host city and country government to guarantee that they will cover the cost overruns of the Games. Our data suggest the guarantee is akin to writing a blank check for the event, with certainty that the cost will be more than what has been quoted. In practice, the bid budget is really more of a down payment than it is a budget, with further installments to be paid later.

2. The Olympic Games have *the highest average cost overrun of any type of megaproject*, at 156 percent in real terms. In comparison, Flyvbjerg et al (2002) found average cost overruns in major transportation projects of 20 percent for roads, 34 percent for large bridges and tunnels, and 45 percent for rail; Ansar et al. (2014) found 90 percent overrun for megadams; and Budzier and Flyvbjerg (2011) 107 percent for major IT projects, all in real terms (see Table 4). The high cost overrun for the Games may be related to the fixed deadline for project delivery: the opening date cannot be moved. Therefore, when problems arise there can be no trade-off between schedule and cost, as is common for other megaprojects. All that managers can do at the Olympics is throw more money at problems, which is what happens. This is the blank check, again.

3. The high average cost overrun for the Games, combined with the existence of outliers, should be cause for caution for anyone considering hosting the Games, and especially small or fragile economies with little capacity to absorb escalating costs and related debts. Even a small risk of a 50+ percent cost overrun on a multi-billion dollar project should concern government officials and taxpayers when a guarantee to cover cost overrun is imposed, because such overrun may have fiscal implications for decades to come, as happened with Montreal where it took 30 years to pay off the debt incurred by the 720 percent cost overrun on the 1976 Summer Games (Vigor, Mean, and Tims 2004: 18), and Athens 2004 where Olympic cost overruns and related debt exacerbated the 2007-16 financial and economic crises, as mentioned above (Flyvbjerg 2011).

*Table 4: The Olympic Games have the largest cost overrun of any type of large-scale project, real terms*

|  | **Roads** | **Bridges, tunnels** | **Energy** | **Rail** | **Dams** | **IT** | **Olympics** |
|---|---|---|---|---|---|---|---|
| **Cost overrun** | 20% | 34% | 36% | 45% | 90% | 107% | 156% |
| **Frequency of cost overrun** | 9 of 10 | 9 of 10 | 6 of 10 | 9 of 10 | 7 of 10 | 5 of 10 | 10 of 10 |
| **Schedule overrun** | 38% | 23% | 38% | 45% | 44% | 37% | 0% |
| **Schedule length, years** | 5.5 | 8.0 | 5.3 | 7.8 | 8.2 | 3.3 | 7.0 |

## Does the Olympic Games Knowledge Management Program Work? Has Cost Overrun Come Down Over Time?

If, perversely, one would want to make it as difficult as possible to deliver a megaproject to budget, then one would (1) make sure that those responsible for delivering the project had never delivered this type of project before, (2) place the project in a location that had never seen such a project, or at least not for the past few decades so that any lessons learned earlier would have been forgotten, and (3) enforce a non-transparent and corrupt bidding process that would encourage overbidding and "winner's curse" and place zero responsibility for costs with the entity that would decide who wins the bid.[5] This, unfortunately, is a fairly accurate description of the playbook for the Olympic Games, as they move from nation to nation and city to city, forcing hosts into a role of "eternal beginners." It is also a further explanation of why the Games hold the record for the highest cost overrun of any type of megaproject, as shown above.[6]

During the 1990's, the IOC began to see that more effective knowledge transfer between host cities might be a way to counter the "eternal beginner" syndrome. The Committee therefore initiated what has become known as the Olympic Games Knowledge Management Program, which is a knowledge-transfer program aimed at increasing efficiency in delivering the Games by having new host cities and nations learn from earlier ones. The key ingredients in the program are, first, a platform of relevant, accumulated knowledge and services that hosts can draw on and, second, a program to have people who will be responsible for future Games participate as trainees and observers at previous ones.

---

[5] The "winner's curse" says that in auctions with incomplete information the winner will tend to overpay and will therefore be worse off than anticipated by winning the bid. On corruption in the Games, see Transparency International (2016).

[6] Recently, proposals have been made to host the Olympic Games in one or a few permanent locations, or, alternatively, that two successive Games should be given to the same host, so facilities could be used twice (Short 2015, Baade and Matheson 2016).

Like for knowledge-transfer programs in general, the purpose of the Olympic Games Knowledge Management Program is to facilitate and support organizational learning. As observed by Schön (1994: 69), learning can only be said to take place if performance improves over time. "Performance that deteriorates, regresses, or merely swings from one mode of action to another does not qualify as learning," according to Schön. In the context of cost, improved performance would mean a reduction in cost risk – i.e., cost overrun – over time. If the data show such reduction, they support that learning is taking place and that the Olympic Games Knowledge Management Program therefore works in this respect. If the data show no reduction in overruns, one would have to conclude that no learning is taking place and that the Program therefore does not work, as regards cost and cost overrun.

The Olympic Games Knowledge Management Program was first used from the mid 1990's in preparing for the Sydney 2000 Summer Games and has been used for Games since then. We therefore decided to compare cost overruns for the Games before 1999 (i.e., before Sydney 2000) with overruns after 1999 (i.e., from, and including, Sydney 2000). For the first few years after 1999 it was impossible to do this comparison in a statistically valid manner because the Games happen only every two years and as a consequence too few observations existed. This situation has now improved with nine Games with valid data on cost overrun after 1999, compared with ten before 1999.

Table 5 shows these data. Eyeballing the table we see that there appears to be a difference in cost overruns for the Games before and after 1999. Median cost overrun for Games before 1999 was 166 percent, compared with 51 percent after 1999. Statistical test shows the difference to be statistically significant (W = 76, p = 0.0101). Cost risks appear to have come down after the Olympic Games Knowledge Management Program began. It should be emphasized, however, that the number of observations are few and variation large. The result of the statistical analysis is therefore sensitive to even small changes in the basic data. For instance, just one or two future Games with triple-digit cost overruns would render the result insignificant and the Knowledge Management Program ineffective, from a statistical point of view. The IOC would therefore be well advised to ensure that the Program is rigorously enforced going forward and to not allow repetition of outlier cost overruns like that of Sochi 2014.

*Table 5: Sports-related cost overrun in the Games pre-1999 and post-1999, real terms. The difference is statistically significant (p = 0.0101).*

| Pre-1999 Games | % cost overrun | Post-1999 Games | % cost overrun |
|---|---|---|---|
| Grenoble 1968 | 181 | Sydney 2000 | 90 |
| Montreal 1976 | 720 | Salt Lake City 2002 | 24 |
| Lake Placid 1980 | 324 | Athens 2004 | 49 |
| Sarajevo 1984 | 118 | Torino 2006 | 80 |
| Calgary 1988 | 65 | Beijing 2008 | 2 |
| Albertville 1992 | 137 | Vancouver 2010 | 13 |
| Barcelona 1992 | 266 | London 2012 | 76 |
| Lillehammer 1994 | 277 | Sochi 2014 | 289 |
| Atlanta 1996 | 151 | Rio 2016* | 51 |
| Nagano 1998 | 56 | - | - |
| *Average* | *230* | *Average* | *75* |
| *Median* | *166* | *Median* | *51* |

*Sources are listed under References. *) Projected final Rio 2016 costs have been used.*

Figure 3 shows how cost overrun has developed over time. It appears there was a general trend of falling cost overruns even before the Sydney 2000 Games and that this continued until Beijing 2008 and Vancouver 2010, after which the trend was reversed, first by London 2012 and then further reinforced by the high overrun for Sochi 2014.

*Figure 3: Development of cost overrun at the Olympic Games over time, real terms*

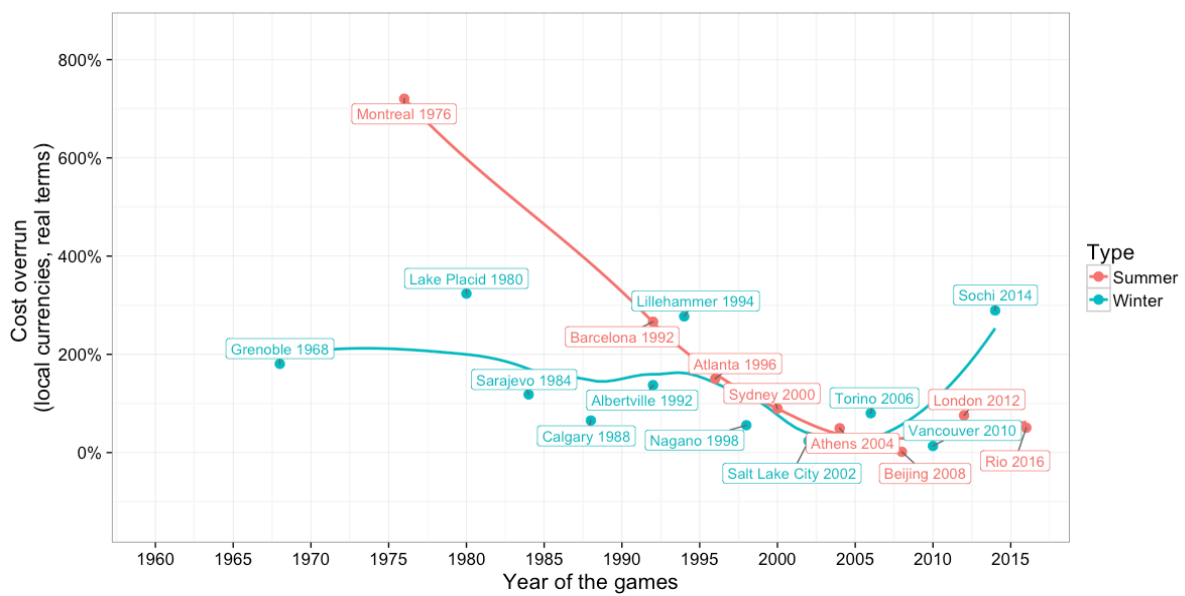

## Rio 2016 Compared with Previous Games

The Rio 2016 Summer Games had not yet been held at the time of writing the present paper, as mentioned above. The latest projected cost, from January 2016, are therefore used for Rio, until such a time when final outturn cost will be available. Using these data, we find a preliminary cost for the Rio Games at USD 4.6 billion and a cost overrun of 51 percent in real terms.

Table 6 compares the Rio 2016 Olympic Summer Games with previous Summer Games for the four key cost variables considered in this paper.

*Table 6: Cost and cost overrun for Rio 2016 compared with previous Summer Games*

| Item | Rio 2016* | Median, Summer Games** |
|---|---|---|
| Cost, billion USD 2015 | 4.6 | 5.0 |
| Cost overrun, % real terms | 50.6 | 89.7 |
| Cost per event, mio. USD 2015 | 14.9 | 16.8 |
| Cost per athlete, mio. USD 2015 | 0.4 | 0.5 |

*) The current projected figures for the Rio 2016 Games have been used.  **) Excluding Rio 2016.

We see that for total sports-related outturn cost, the Rio 2016 Games are just below the median cost of all other Summer Games, which is USD 5.0 billion. The difference is non-significant, however (V = 37, p = 0.7646). We therefore conclude that the Rio 2016 Games are similar to other Summer Games in terms of cost.

For cost overrun, at 51 percent the Rio Games are lower than the median for other Summer Games, but again the difference is not statistically significant (V = 23, p = 0.156). Rio's overrun is identical to median overrun for Games since 1999 (see Table 5), but is less than the overruns for the most recent Games in London and Sochi.

Measured by cost per event, the Rio 2016 Games – at USD 14.9 million – are slightly lower than the median cost per event of previous Summer Games at USD 16.8 million. But cost per event for Rio 2016 is not statistically significantly different than cost per event for previous Summer Games (V = 45, p = 0.3203).

Finally, for cost per athlete, the Rio Games – at USD 0.4 million – are again slightly lower than the cost per athlete of previous Summer Games at USD 0.5 million. And again the difference is statistically non-significant (V = 44, p = 0.3652).

In sum, we find for the Rio 2016 Olympic Summer Games, based on the current projected cost figures for Rio:

1. At USD 4.6 billion, the Rio Games appear to be on track to reverse the high expenditures of London 2012 and Sochi 2014 and deliver a Summer Games at the median cost for such Games.
2. At 51 percent in real terms, Rio 2016 seems on track to incur a substantial cost overrun of USD 1.6 billion. This is less than the overruns for the most recent Games in London and Sochi, but the same as median overrun for Games since 1999.
3. At USD 0.4 million per athlete, Rio is on track to deliver its Games at a cost per athlete similar to previous Summer Games and substantially lower than both London 2012 and Sochi 2014.

**Appendix: Research Methodology**

To investigate the question of cost and cost overrun at the Olympic Games, we conducted an extended search of all cost data available. We collected data on two cost components for the Games: the OCOG cost and non-OCOG direct costs. For these two components, costs at bid were determined mainly through primary data from the OCOG Candidature Files. Final OCOG costs were found chiefly in the Official Reports submitted to the IOC following each Games. Where primary sources were not available, secondary sources including audits and other research data were used, with primary sources taking precedence over secondary sources where available.

Using these data, we then proceeded with cost comparisons. For calculating cost overrun, we compared the actual outturn cost with the estimated cost in the bid year, both listed in local currency and calculated in real terms. To control for inflation over the period from bid to reporting of final outturn costs, we adjusted the cost data using local GDP deflator values to adjust the cost in local currency to the bid year, using a distribution of costs over the seven years of Games planning based on known expenditure profiles of OCOG costs and non-OCOG direct costs. The numbers for cost overrun arrived at in this manner are conservative, in the sense that they would be substantially larger had inflation been included. Finally, for cost comparisons in 2015 US dollars, we used the same local GDP deflator values (World Bank 2016a) to inflate the nominal bid and final cost data in the year in which it was incurred to 2015 in local currency, and then used World Bank national currency unit (NCU) values (World Bank 2016b) to convert the data from 2015 local currency to USD 2015.

We next conducted a number of statistical tests to understand the differences present in the data. Table 7 shows the hypotheses tested, the methods used, and the results found.

*Table 7: Hypotheses tested, methods used, and results found in the Oxford Olympics Study*

| Hypothesis | Data | Test used | Result |
| --- | --- | --- | --- |
| Median cost are different between Summer and Winter Olympics | Actual cost in 2015 USD terms | Wilcoxon sum rank test | W = 38, p = 0. 02982 |
| Cost of Summer Olympics are stationary | Actual cost in 2015 USD terms | Augmented Dickey-Fuller Test | Dickey-Fuller = -2.3284, p = 0.4473 |
| Cost of Winter Olympics are stationary | Actual cost in 2015 USD terms | Augmented Dickey-Fuller Test | Dickey-Fuller = -1.4969, p = 0.764 |
| Cost per athlete are different between Summer and Winter Olympics | Actual cost in 2015 USD terms per athlete | Wilcoxon sum rank test | W = 97, P = 0.3203 |
| Cost per athlete of Summer Olympics are stationary | Actual cost in 2015 USD terms per athlete | Augmented Dickey-Fuller Test | Dickey-Fuller = -2.2525, p = 0.4762 |
| Cost per athlete of Winter Olympics are stationary | Actual cost in 2015 USD terms per athlete | Augmented Dickey-Fuller Test | Dickey-Fuller = -3.1027, p = 0.1523 |
| Cost per athlete are stationary | Actual cost in 2015 USD terms per athlete | Augmented Dickey-Fuller Test | Dickey-Fuller = -2.0852, p = 0.5399 |
| Cost overruns center around 0% | Actual/estimated cost, local currencies, real terms | Wilcoxon signed rank test | V = 190, p < 0.0001 |
| Cost overruns are no different between Summer and Winter Olympics | Actual/estimated cost, local currencies, real terms | Wilcoxon sum rank test | W = 49, p = 0.7168 |
| Cost overruns are no different post/pre-1999 | Actual/estimated cost, local currencies, real terms | Wilcoxon sum rank test | W = 76, p = 0.01013 |
| Rio 2016 outturn cost are similar to previous Summer Olympics | Actual cost, 2015 USD, real terms | Wilcoxon sum rank test | V = 37, p = 0.7646 |
| Rio 2016 outturn cost are similar to previous Olympics | Actual cost, 2015 USD, real terms | Wilcoxon sum rank test | V = 91, p = 0.0951 |
| Rio 2016 cost overrun is similar to previous Summer Olympics | Actual/estimated cost, local currencies, real terms | Wilcoxon sum rank test | V = 23, p = 0.1563 |
| Rio 2016 cost overrun is similar to previous Olympics | Actual/estimated cost, local currencies, real terms | Wilcoxon sum rank test | V = 149, p = 0.004005 |
| Rio 2016 cost per event are similar to previous Summer Olympics | Actual cost per event, 2015 USD, real terms | Wilcoxon sum rank test | V = 45, p = 0.3203 |
| Rio 2016 cost per athlete are similar to previous Summer Olympics | Actual/estimated cost, local currencies, real terms | Wilcoxon sum rank test | V = 44, p = 0.3652 |

**References**

Albertville Bid Committee (1985). Candidature File. Accessed at the Olympic Studies Centre (OSC): Lausanne, Switzerland, September 2011.

Albertville Organising Committee for the Olympic Games (1992). Official Report. Available for download from the LA84 Foundation at http://www.la84foundation.org/.

Andranovich, G., M. J. Burbank, and C. H. Heying (2001) Olympic Cities: Lessons Learned from Mega-event politics. Journal of Urban Affairs, 23(2): 113-131.

Ansar, A., Flyvbjerg, B., Budzier, A., and Lunn D. (forthcoming). "Big Is Fragile: An Attempt at Theorizing Scale", in Bent Flyvbjerg, ed., The Oxford Handbook of Megaproject Management (Oxford: Oxford University Press).

Anti-Olympics People's Network (1998). The Real Story of the Nagano Olympics. Available at: http://nolimpiadi.8m.com/nagano1.html

Athens Bid Committee (1997). Candidature File, Volume 3. Accessed at the OSC: Lausanne, Switzerland, September 2011.

Athens Organising Committee for the Olympic Games (2005). Official Report. Available for download from the LA84 Foundation at http://www.la84foundation.org/.

Atlanta Bid Committee (1989). Candidature File, Volume 1. Accessed at the OSC: Lausanne, Switzerland, September 2011.

Baade, Robert A. and Victor A. Matheson, 2016, "Going for the Gold: The Economics of the Olympics," Journal of Economic Perspectives, vol. 30, no. 2, Spring, pp. 201-218.

Barcelona Bid Committee (1985). Candidature File, 2nd edition. Accessed at the OSC: Lausanne, Switzerland, September 2011.

Barcelona Organising Committee for the Olympic Games (1992). Official Report, Volume 2. Available for download from the LA84 Foundation at http://www.la84foundation.org/.

BBC, British Broadcasting Corporation (2013). London 2012: Olympics and Paralympics £528m Under Budget, July 19, http://www.bbc.co.uk/sport/olympics/20041426, accessed June 14, 2016.

Beijing Bid Committee (2001). Candidature File. Finance section, p 73. Accessed at the OSC: Lausanne, Switzerland, September 2011.

Bondonio, P. and N. Campaniello (2006). Torino 2006: What Kind of Olympic Winter Games Were They? A Preliminary Account From an Organizational and Economic Perspective. Olympika, 15: 355-380.

Brunet, F. (1995). An Economic Analysis of the Barcelona '92 Olympic Games: Resources, Financing and Impact. Available at: http://olympicstudies.uab.es/pdf/OD006_eng.pdf

Budzier, A. and B. Flyvbjerg (2011). Double Whammy – How ICT Projects are Fooled by Randomness and Screwed by Political Intent. Saïd Business School working papers.

Calgary Bid Committee (1981). Candidature File. Accessed at OSC: Lausanne, Switzerland,  September 2011.

Calgary Organising Committee for the Olympic Games (1989). Official Report, Volume 1. Available for download from the LA84 Foundation at http://www.la84foundation.org/.

CBC News (19 December 2006). Quebec's Big Owe stadium debt is over. Available at: http://www.cbc.ca/news/canada/montreal/story/2006/12/19/qc-olympicstadium.html

Chappelet, J-L. (2002). From Lake Placid to Salt Lake City: The challenging growth of the Olympic Winter Games Since 1980. European Journal of Sport Science, 2(3): 1 - 21.

Comitê Organizador dos Jogos Olímpicos e Paralímpicos Rio 2016 (2016). Rio 2016 Organising Committe - Budget, Available at: https://www.rio2016.com/transparencia/en/budget

Comitê Organizador dos Jogos Olímpicos e Paralímpicos Rio 2016 (March 2016). Financial Statements 2015, available at: https://www.rio2016.com/transparencia/sites/default/files/fnc_demonstracoes_financeiras_2015_2016_ingles_11042016_0.pdf

Commission d'enquete sur le cout de la 21e olympiade (April 1980). Rapport. Accessed at the OSC: Lausanne, Switzerland, 721(086) MA 6754/1. de Coubertin, P., 1911. Olympic review, 6: 59-62.

Department for Culture, Media and Sport (28 February 2012) The London 2012 Olympic and Paralympic Games remain on time and within budget. News Release, available at: http://www.culture.gov.uk/news/media_releases/8891.aspx

Department for Culture, Media and Sport (June 2012). London 2012 Olympic and Paralympic Games Quarterly Report June 2012, London: UK Department for Culture, Media and Sport.

Department for Culture, Media and Sport (October 2012). London 2012 Olympic and Paralympic Games Quarterly Report October 2012, London: UK Department for Culture, Media and Sport.

Essex, S. and B. Chalkley (2004). Mega-sporting events in urban and regional policy: a history of the Winter Olympics. Planning Perspectives, 19 (April): 201–232.

Flyvbjerg, B. (2008). Curbing Optimism Bias and Strategic Misrepresentation in Planning: Reference Class Forecasting in Practice. European Planning Studies, 16(1) Jan 2008.

Flyvbjerg, B. (2011). Over Budget, Over Time, Over and Over Again: Managing Major Projects, in Peter W. G. Morris, Jeffrey K. Pinto, and Jonas Söderlund, eds., The Oxford Handbook of Project Management (Oxford: Oxford University Press), pp. 321-344.

Flyvbjerg, B. (2016). The Fallacy of Beneficial Ignorance: A Test of Hirschman's Hiding Hand, World Development, vol. 84, May, pp. 176–189,

Flyvbjerg, B., Holm, M. K., and Buhl, S. L. (2002). Underestimating costs in public works projects: Error or lie? Journal of the American Planning Association, 68(3), Summer, pp. 279–295.

Flyvbjerg, B. and Stewart, A. (2012). Olympic Proportions: Cost and Cost Overrun at the Olympics 1960-2012. Working Paper, Saïd Business School, University of Oxford, June 2012.

Grenoble Organising Committee for the Olympic Games (1969). Official Report of the Xth Winter Olympic Games, pages 42, 332-334. Available for download from the LA84 Foundation at http://www.la84foundation.org/.

Grenoble Organising Committee for the Olympic Games (1965). Report Presented to the International Olympic Committee on the Occasion of its 63rd session. Madrid, Oct 6-9, 1965. p 9-10. Accessed at the OSC: Lausanne, Switzerland, September 2011.

Innsbruck 1976 Organising Committee (1976). Report to the IOC, 1976, Innsbruck: Innsbruck Organizing Committee.

IOC (1996). The IOC: One hundred years – The idea, the Presidents, The achievements. IOC: Lausanne. Volume III.

IOC (2004). 2012 Candidature Procedure and Questionnaire: Games of the XXX Olympiad in 2012.   Available at:  http://www.olympic.org/Documents/Reports/EN/en_report_810.pdf
IOC (2008) Managing Olympic Knowledge. IOC Focus, December 2008. Available upon request.

IOC (2009). Report of the 2016 IOC Evaluation Commission, available at: https://www.olympic.org/Documents/Reports/EN/en_report_1469.pdf

Jennings, W. (2012). Why costs overrun: risk, optimism and uncertainty in budgeting for the London 2012 Olympic Games. Construction Management and Economics, 2012: 1-8.

Kahneman, D. (2011). Thinking, Fast and Slow. New York: Farrar, Straus and Giroux.

Koch-Weser, Iacob N., (2013). The Reliability of China's Economic Data: An Analysis of National Output, Washington, DC: US-China Economic and Security Review Commission, US Congress.

Lake Placid Bid Committee (1973). Lake Placid, Candidate for 1980 Winter Olympic Games: General Information. Section: Proposed Sources of Funding for Additional Required Facilities. Accessed at the OSC: Lausanne, Switzerland, 722.13 LAK, MA 6850, September 2011.

Lake Placid Organizing Committee for the Olympic Games (1981). Official Report, Volume 1. Available for download from the LA84 Foundation at http://www.la84foundation.org/.

Lillehammer Bid Committee (1987). Candidature File. Accessed at the OSC: Lausanne,  Switzerland, September 2011.

Lillehammer Organising Committee for the Olympic Games (1995). Official Report, Volume 1, p 29. Available for download from the LA84 Foundation at http://www.la84foundation.org/.

LOCOG (2013). London 2012 Olympic Games: The Official Report, Organising Committee for the Olympic and Paralympic Games in London in 2012, London: LOCOG, 2013.

London Bid Committee (2005). London Candidature File. Vol 1. Accessed at the OSC: Lausanne, Switzerland, September 2011.

Los Angeles Organizing Committee (1985). Official Report of the Games of the XXIIIrd Olympiad Los Angeles, 1984, Los Angeles: Los Angeles Organizing Committee.

Malfas, M., E. Theodoraki, and B. Houlihan (2004). Impacts of the Olympic Games as mega-events. Proceedings of the Institution of Civil Engineers, Municipal Engineer, 157(ME3): 209–220.

Mexico 68 (1969). Mexico City Olympic Games Official Report, 1969, Mexico City: Organizing Committee of the Games of the XIX Olympiad.

Moraes, R. (29 January 2016). Cost of Rio's 2016 Olympics rises by almost $100 million, available at: http://www.reuters.com/article/us-olympics-rio-cost-idUSKCN0V726H

Moscow Fizkultura i Sport (1981). Official Report of the Organising Committee of the Games of the XXII Olympiad, Moscow, 1980, Moscow: Moscow Organizing Committee.

Müller, M. (2014). After Sochi 2014: Costs and Impacts of Russia's Olympic Games. Eurasian Geography and Economics, 55(6), 628-655.

Nagano Bid Committee (1991). Candidature File, Volume 2 "Games Organization".
Accessed at the OSC: Lausanne, Switzerland, CIO MB 30/3, 726.18 NAG, September 2011.

Nagano Organising Committee for the Olympic Games (1998). Official Report, Volume 1. Available from LA84 Foundation.

National Audit Office of the People's Republic of China (19 June 2009). No. 8 of 2009 (Serial No. 40): Follow-up Audit Findings on the Revenues and Expenditures of the Beijing Olympics and the Construction of Olympics Venues. Available at: http://www.cnao.gov.cn/main/articleshow_ArtID_1057.htm.

New South Wales Audit Office (1998). Performance Audit Report: The Sydney 2000 Olympic and Paralympic Games, Review of Estimates. Available at:
http://auditofficestaging.elcom.com.au/ArticleDocuments/131/62_Sydney_Olympic_Games.pdf.aspx?Embed=Y.

Neumann, B. and Organisationskomitee für die Olympischen Winterspiele in Innsbruck in 1976 (1976). Endbericht, 1976, Innsbruck: Organisationskomitee der XII. Olympischen Winterspiele.

Organizing Committee of the Games of the XVII Olympiad (1960). The XVII Olympiad Rome 1960: The Official Report of the Organizing Committee, 1960, Rome: Organizing Committee of the Games of the XVII Olympiad.

Organizing Committee for the Games of the XVIII Olympiad (1964). The Games of the XVIII Olympiad Tokyo 1964: The Official Report of the Organizing Committee, 1964, Tokyo: Organizing Committee for the Games of the XVIII Olympiad.

Organising Committee for the Olympic Winter Games in Sapporo in 1972 (1973). The XI Olympic Winter Games Sapporo 1972: Official Report, 1973, Sapporo: Le Comité d'organisation des XIèmes Jeux Olympiques d'hiver.

Personal Correspondence (16 April 2012). Email from Terry Wright, Vancouver OC.

Preuss, H. (2004). The economics of staging the Olympics: a comparison of the Games, 1972-2008. Cheltenham: Edward Elgar Publishing Ltd.

Pro Sport München (1973). The official report of the Organizing Committee for the Games of the XXth Olympiad Munich 1972, 1973, Munich: pro Sport.

Rubin, R. and Organizing Committee for the Olympic Winter Games in Squaw Valley in 1960 (1960). VIII Olympic Winter Games Squaw Valley, California, 1960: final report, 1960, Squaw Valley: Organizing Committee of the Games.

Salt Lake City Bid Committee (1995). Candidature File, Volume 3, p 72-74. Accessed at the OSC: Lausanne, Switzerland, September 2011.

Salt Lake City Organizing Committee for the Olympic Games (2003). Official Report, Volume 1. Available for download from the LA84 Foundation at http://www.la84foundation.org/.

Sarajevo, Bosnia, and Herzegovina, Yugoslavia (1977). Technical Questionnaire, Finances section. Accessed at the OSC: Lausanne, Switzerland, 723.14 SAR, MA 7086, September 2011.

Sarajevo Organising Committee for the Olympic Games (1985). Official Report, p 184. Available for download from the LA84 Foundation at    http://www.la84foundation.org/.

Seoul Organising Committee (1983). Games Report. Seoul: Organising Committee, 1983.

Preuss, H. (2006). Lasting Effects of Major Sporting Events. Idrottsforum, (4), 1–6.

Short, J. R. (2015). "We Should Host the Olympics in the Same Place Every Time," The Washington Post, July 28.

Sprent, P. (1989). Applied nonparametric statistical methods. London: Chapman and Hall.

Stewart, A. (forthcoming). DPhil Dissertation. University of Oxford, Saïd Business School, available January 2013.

Sydney Bid Committee (1993). Candidature File, Volume 3, p 38. Accessed at the OSC: Lausanne, Switzerland, September 2011.

Sydney Organising Committee for the Olympic Games (2001). Official Report, Volume 1. Available for download from the LA84 Foundation at http://www.la84foundation.org/.

Torino Bid Committee (1999). Candidature File. Volume 1. Accessed at the OSC: Lausanne, Switzerland, September 2011.

Torino Organising Committee for the Olympic Games (2007). Official Report. Vol 2. Available for download from the LA84 Foundation at http://www.la84foundation.org/.

Transparency International (2016). Global Corruption Report: Sport, London and New York: Routledge.

UK Public Accounts Committee (14 December 2011). House of Commons Oral Evidence Taken Before the Public Accounts Committee Management Preparations for the Olympics. Corrected Transcript of Oral Evidence To be published as HC 1716-i. Available at:
http://www.publications.parliament.uk/pa/cm201012/cmselect/cmpubacc/c1716- i/c171601.htm

UK Public Accounts Committee (29 February 2012). Seventy-Fourth Report: Preparations for the London 2012 Olympic and Paralympic Games. Available at:
http://www.publications.parliament.uk/pa/cm201012/cmselect/cmpubacc/1716/171602. htm

Vancouver Bid Committee (2003). Candidature File. Finance and Administration: 81. Accessed at the OSC: Lausanne, Switzerland, September 2011.

Vigor, A., M. Mean, and C. Tims (2004). After the gold rush: a sustainable Olympics for London. London: IPPR.

Wolfgang, F. and Organisationskomitee für die Olympischen Winterspiele in Innsbruck in 1964 (1967). Offizieller Bericht der IX. Olympischen Winterspiele Innsbruck 1964, 1967, Wien: Österreichischer Bundesverl. für Unterricht-Wissenschaft und Kunst.

World Bank (1 June, 2016a). GDP Deflator (base year varies by country). Available at:
http://data.worldbank.org/indicator/NY.GDP.DEFL.ZS

World Bank (1 June, 2016b). Official exchange rate (LCU per US$, period average). Available at:
http://data.worldbank.org/indicator/PA.NUS.FCRF

Zimbalist, A. (2015). Circus Maximus: The Economic Gamble Behind Hosting the Olympics and the World Cup, Washington, DC: Brookings Institution Press.